\documentclass[12pt]{article}
\usepackage{amsmath}
\usepackage{amssymb}
\usepackage{graphicx}
\usepackage{enumerate}
\usepackage{natbib}
\usepackage{url} 

\newcommand{\blind}{1}

\newtheorem{theorem}{Theorem}
\newtheorem{corollary}{Corollary}

\newcommand{\abs}[1]{\left\vert#1\right\vert}
\newcommand{\E}{\mathbb{E}}
\newcommand{\R}{\mathbb R}

\begin{document}

\def\spacingset#1{\renewcommand{\baselinestretch}%
{#1}\small\normalsize} \spacingset{1}


\if1\blind
{
  \title{\bf R.~A.~Fisher's Exact Test Revisited}
  \author{Martin Mugnier\thanks{
     Mailing address: 10 Manor Road, Oxford, OX1 3UQ, United Kingdom. Contact: martin.mugnier@economics.ox.ac.uk. The author gratefully acknowledges St\'{e}phane Bonhomme for his comments. 
    }\hspace{.2cm}\\
    Department of Economics, University of Oxford
    }
  \maketitle
} \fi

\if0\blind
{
  \bigskip
  \bigskip
  \bigskip
  \begin{center}
    {\LARGE\bf R.~A.~Fisher's Exact Test Revisited}
\end{center}
  \medskip
} \fi

\bigskip
\begin{abstract}
This note provides a conceptual clarification of Ronald Aylmer~Fisher's (1935) pioneering exact test in the context of the Lady Testing Tea experiment. It unveils a critical implicit assumption in Fisher’s calibration: the taster minimizes expected misclassification given fixed probabilistic information. Without similar assumptions or an explicit alternative hypothesis, the rationale behind Fisher’s specification of the rejection region remains unclear. 
\end{abstract}

\noindent%
\begin{center}
\end{center}
\vfill

\newpage
\spacingset{1.9} 
\section{Introduction}
\label{sec:intro}

A colleague of Ronald Aylmer Fisher asserts that they can distinguish tea poured into milk from milk poured into tea. To convince himself, Fisher prepares eight cups of tea, four with tea poured into milk (confidentially labelled $TM$) and four with milk poured into tea (confidentially labelled $MT$). He presents the cups in random order to his colleague, who is asked to taste and assign one of the labels $\{TM,MT\}$ to each cup. Before tasting, the colleague knows that there are four $TM$ cups and four $MT$ cups among the eight, but they do not know the specific order in which Fisher arranged the cups.\footnote{I am indebted to \cite{ritzwoller2024} for inspiring this short paper and introduction. This experiment is reported in \cite{Fisher1935}, p.13-29.} 

\section{A disconnect between Fisher's null hypothesis and exact test}
\label{sec:disconnect}

\paragraph{Fisher's null hypothesis} Distinguishing $TM$ cups from $MT$ cups is a rather surprising skill. Fisher considers the null hypothesis that all possible assignments of eight cups into two groups of four are equally probable answers. He disbelieves that his colleague can do better or worse than a fair coin toss prediction unless they provide some experimental evidence. Specifically, Fisher states
\begin{quote}
The two classes of results which are distinguished by our test of significance are, on the one hand, those which show a significant discrepancy from a certain hypothesis; namely, in this case, the hypothesis that the judgments given are in no way influenced by the order in which the ingredients have been added; and on the other hand, results which show no significant discrepancy from this hypothesis. \citep[][p.18]{Fisher1935}
\end{quote}
Without further assumption, experimental evidence can be understood as observing some event reflecting a ``sufficiently large deviation'' from what the uniform distribution over the set of all possible assignments would ``typically predict''. Such an event may be {\em perfect distinction in the prediction sense}, which occurs if each time the experimenter presents $TM$ (resp.~$MT$), the taster claims $TM$ (resp.~$MT$). Another one is {\em perfect distinction in the weak information sense}, which occurs if each time the experimenter presents $TM$ (resp.~$MT$), the taster claims $MT$ (resp.~$TM$).  Note that perfect distinction in the weak informative sense implies perfect distinction in the predictive sense up to applying the deterministic bijective transformation $g:\{TM,MT\}\to\{TM,MT\}$ defined as $g(TM)=MT$ and $g(MT)=TM$ to each claim. In other words, both extreme events equally suggest that the taster can ``distinguish'' by using some knowledge about $TM$ and $MT$ to recover the correct equivalence classes of cups invariant to relabeling. 

\paragraph{Fisher's exact test}
Let $Y_N=(Y_{1N},\ldots,Y_{NN})'$ denote the $N$ cup labels corresponding to the random ordering chosen by Fisher, among which $n$ are $TM$ and $N-n$ are $MT$. Let $X_N=(X_{1N},\ldots,X_{NN})'$ denote the (random) sequence of answers from his colleague. Fisher's null hypothesis is an assumption on the joint distribution $P_{(X_N,Y_N)}$. Specifically, it says that the conditional distribution $P_{X_N|Y_N}$ is the uniform distribution over the set 
$$\mathcal X_{N,n}:=\left\{x\in\{TM,MT\}^N:\sum_{j=1}^N1\{x_j=TM\}=n \right\}.$$
In Fisher's experiment, $N=8$, $n=4$, and $\vert\mathcal X_{8,4}\vert={8\choose 4}=70$ where for any finite set $A$, $\vert A \vert$ denotes the cardinality of $A$. Fisher randomly draws $Y_8\in\mathcal X_{8,4}$ and his colleague answers $X_8\in\mathcal X_{8,4}$. Under the null hypothesis,  for any given $Y_8\in\mathcal X_{8,4}$, there is a 1/70 chance of observing only right answers, i.e., $X_8=Y_8$. Fisher rejects the null hypothesis if and only if $X_8=Y_8$.  The test is exact at the significance level $5\%$ since the probability to reject under the null, i.e., to make a Type I error, is $1/70\approx0.014\leq5\%$. This probability does not depend on $Y_8$: it is the same both conditionally and unconditionally on $Y_8$.

\paragraph{A disconnect} 
Though Fisher does not specify an alternative hypothesis, he chooses a rejection region in favour of perfect distinction in the predictive sense. Why ignore the region in favour of perfect distinction in the weak information sense? Certainly not for significance-level considerations. Under the null,  the event of observing no right answer, $\cap_{j=1}^8\{X_{j8}\neq Y_{j8}\}$, also occurs with probability 1/70 so that the exact test that rejects the null if and only if $\cap_{j=1}^8\{X_{j8}\neq Y_{j8}\} \cup \{X_8=Y_8\}$ is observed is at least as powerful against all alternatives, is more powerful against some alternatives, and enjoys the same significance level since $2/70\approx0.029\leq5\%$.  

In this experiment, the colleague ends up making correct predictions for all cups, which implies that both tests reject the null at the significance level $5\%$ (the size of the more powerful test, however, is higher). Despite that running the more powerful test does not change the conclusion in this particular experiment, there is a priori no theoretical justification for preferring the original exact test. More concerning, the exact $5\%$-level test that rejects the null if and only if $\cap_{j=1}^8\{X_{j8}\neq Y_{j8}\}$ is observed, although equally legitimate, does not reject the null in this experiment.  

Fisher is one of the most celebrated statisticians and it is of interest to investigate what could have motivated the disconnect between his null hypothesis and his proposed exact test. A natural justification is that Fisher has an alternative hypothesis in mind. Others have gathered around the table, and the only reason his colleague might challenge him is to demonstrate to the community that they can do ``better'' than a random guess in the commonly accepted sense, i.e., the prediction sense. Fisher is willing to reject only based on evidence pointing to that direction. This justification, however, is rather unsatisfactory for at least two reasons. First, \cite{Neyman1928} are credited for introducing the concept of alternative hypothesis after Fisher's first formulation of his methodology \citep{Fisher1925}. Second, Fisher himself later opposed the use of an alternative hypothesis in the 1966 edition of his book \citep{Fisher1935}.

The next section provides another rigorous justification taking the form of an explicit behavioural hypothesis on the taster that avoids explicitly invoking any particular alternative hypothesis. It amounts to formalizing a single crucially important sentence in the original text, though left over in the sequel and often overlooked in modern expositions: 
\begin{quote}
	Her task [the taster] is to divide the 8 cups into two sets of 4, {\em agreeing, if possible, with the treatments received}.\footnote{Italics and text into square brackets added.} \cite[][p.14]{Fisher1935}
\end{quote}
Fisher implicitly assumes that he and the taster agree that the taster will use their discriminating ability,  {\em if any} (that is {\em under any alternative}), to minimize misclassification. While this assumption is innocuous to stating the null hypothesis, it is central to justifying Fisher’s test rejection region, as it implies stochastic dominance relationships between distributions of success under different levels of probabilistic information. Theorem 3.1 clarifies this point by formalizing the minimization problem within a statistical decision and information-theoretic framework. 

\section{A reconciling statistical decision and information-theoretic framework}\label{sec:stat_deci}
As emphasized in the previous section, a possible source of confusion regarding the mapping from Fisher's null hypothesis to his exact test is that the ability to distinguish has more to do with information than correct prediction. 
More than 70 years ago (and thus posterior to Fisher), Claude Shannon introduced the concept of {\em entropy} or {\em Shannon information}.\footnote{See \cite{Shannon1948}.}  The main contribution of this note is to employ this key concept from information theory to eliminate any disconnect by giving a mathematically precise, behaviourally sound and rigorous justification for \cite{Fisher1935}'s exact test in the {\em Lady Testing Tea} experiment. 

\paragraph{Misclassification loss} Consider the classical binary misclassification loss $\ell$, popular in the statistical learning and decision literature: for all $x_N,y_N\in\mathcal X_{N,n}$,
\[
\ell(x_N,y_N):=\sum_{j=1}^N1\{x_{jN}\neq y_{jN}\}.
\]
The theoretical expected predictive performance of the taster in terms of loss $\ell$ and distribution of answers $P_{X_N|Y_N}$ can be visualized as an arrow ``indexing the joint distributions'' from the left (systematically less correct than random with the extremity being ``always wrong'' if almost surely $\cap_{j=1}^8\{X_{j8}\neq Y_{j8}\}$ and thus $\ell(X_8,Y_8)=8$) through the centre (as correct as random) to the right (systematically more correct than random with the extremity being ``always correct'' if almost surely $X_8=Y_8$ and thus $\ell(X_8,Y_8)=0$). Clearly, observing $\cap_{j=1}^8\{X_{j8}\neq Y_{j8}\}$ provides evidence against the null {\em in the direction to the left} (previously called ``information sense'') while observing $X_8=Y_8$ provides evidence against the null {\em in the direction to the right} (previously called ``prediction sense''). 

\begin{figure}[htbp]
    \centering
    \includegraphics[width=0.8\textwidth]{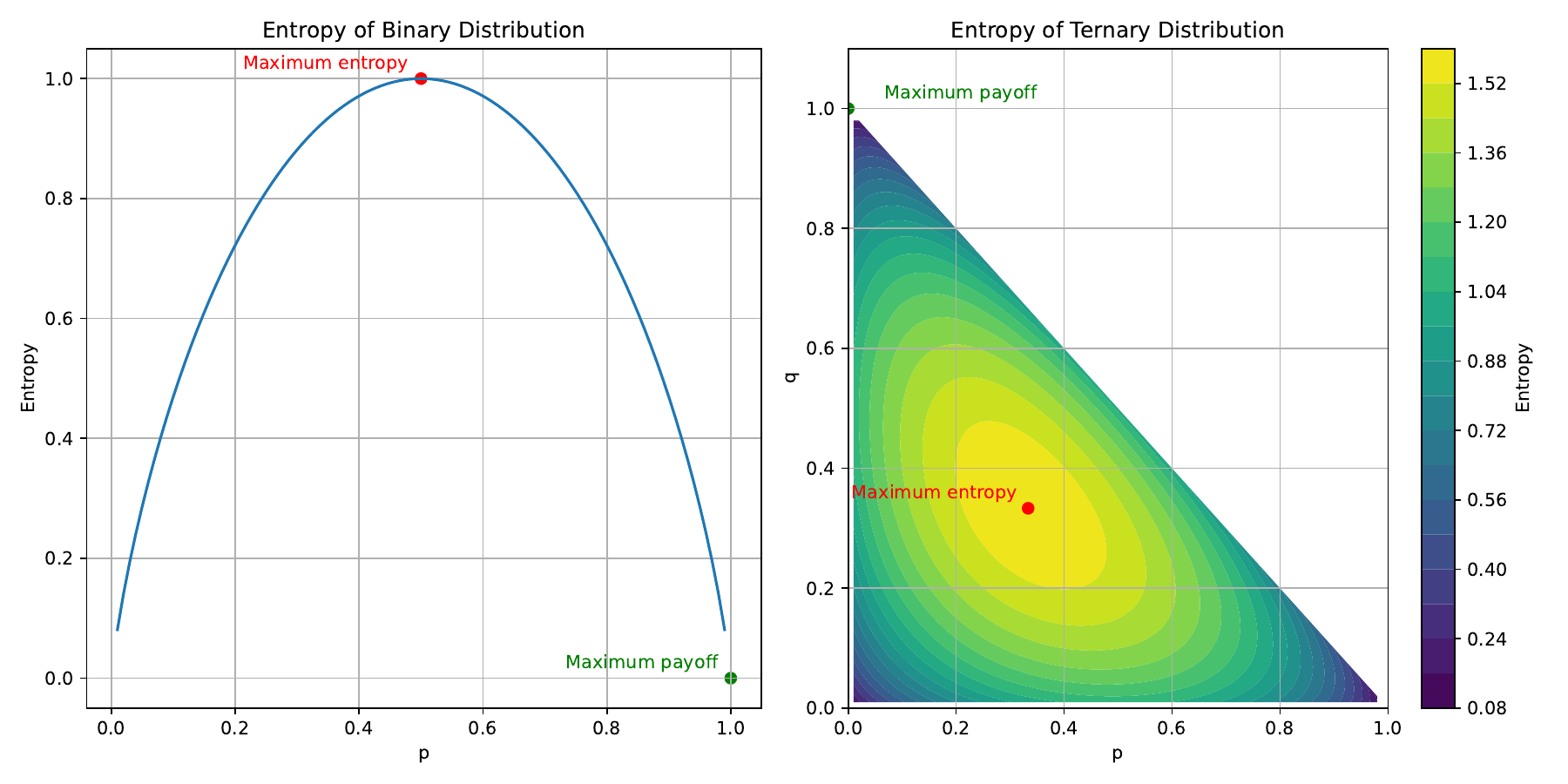}
    \caption{Entropy for binary or ternary random variables with arbitrary maximum payoff points.
    \label{fig:entropy}}
\end{figure}

\paragraph{Misclassification minimization subject to an information constraint} For any distribution $P$ supported on a discrete set $\mathcal X$, define the probabilistic entropy, or Shannon information, as
\begin{equation*}
    H(P)=-\sum_{x\in\mathcal X}P(x)\log(P(x)).
\end{equation*}
Note that $H(X)\in[0,\log(\vert \mathcal X\vert )]$ is uniquely maximized at the uniform distribution $\overline P$ on $\mathcal X$, and minimized for Dirac distributions $\underline P\in \mathcal D(\mathcal X):=\{\delta_x:x\in\mathcal X\}$. These facts are useful to show the main result of the paper stated below.
\begin{theorem}\label{thm:shanon_inf}
    Let $\overline h:=\log(\vert\mathcal X_{N,n}\vert)$. For $h\in(0,\overline h]$, let $P_{X_N|Y_N}^*(h)$ denote an interior solution to the minimization problem
    \[
    \begin{array}{ll}
         & \min_{P_{X_N|Y_N}}\mathbb E_{P_{X_N|Y_N}}[\ell(X_N,Y_N)]  \\
         & \text{ subject to } \quad H(P_{X_N|Y_N})\geq h.
    \end{array}
    \]
    Then,  $P_{X_N|Y_N}^*(h)$ is uniquely defined and such that 
    \begin{enumerate}
        \item $H(P^*_{X_N|Y_N}(h))=h$, \label{res:thm1}
        \item $\lim_{h\to0^+}\sup_{x\in\mathcal X_{N,n}}\vert P_{X_N|Y_N}^*(h)(x)-\delta_{Y_N}(x)\vert=0$, \label{res:thm2}
        \item $P_{X_N|Y_N}^*(\overline h)={\rm Uniform}(\mathcal X_{N,n})$, \label{res:thm3}
        \item For all $h,h'\in(0,\overline h]$ such that $h\leq h'$, $\ell(X_N,Y_N)$ under $P_{X_N|Y_N}^*(h')$ first-order stochastically dominates $\ell(X_N,Y_N)$ under $P_{X_N|Y_N}^*(h)$. \label{res:thm4}
     \end{enumerate}
\end{theorem}
Points \ref{res:thm1}-\ref{res:thm3} of Theorem~\ref{thm:shanon_inf} say that if the taster chooses their answers ($P_{X_N|Y_N}^*(h)$) to minimize the expected misclassification error $\mathbb E_{P_{X_N|Y_N}}[\ell(X_N,Y_N)]$ given a maximum level of information $\overline h-h$, then i) the taster uses all available information, ii) the taster with full information ($h=0^+$) will almost surely give all answers right, and iii) Fisher's null hypothesis corresponds to a taster with no information ($h=\overline h$). Point~\ref{res:thm4} is the most enlightening result since it justifies Fisher's exact test irrelevant of {\em any} alternative. Specifically, the more the taster deviates from the null hypothesis of no information ($h=\overline h$), the more their answers ($P_{X_N|Y_N}^*(h)$) generate a distribution of successes ($N-\ell(X_N,Y_N)$) that has a heavy right tail. Such a minimizing behaviour disciplines the taster's response as a function of their information and solves the previous inconsistency of ``two-sided'' testing with a one-sided test by selecting a unique direction. Corollary~\ref{cor:fisher} displays the main implications of this result for Fisher's exact test. Figure~\ref{fig:entropy} provides an illustration for binary and ternary outcome distributions. The optimal path as a function of $h$ can be traced out starting from the red dot (maximum entropy distribution) and minimizing the distance to the green dot (maximum payoff) along the level curves. Analogous results hold for a maximizing taster.
\begin{corollary}\label{cor:fisher} ~
Suppose the taster minimizes expected misclassification given a maximum probabilistic information level. Then,
	\begin{enumerate}
	\item For all $x,y\in\mathcal X_{8,4}$,
	\begin{align*}
	& \ell(x,y)\leq  \mathbb E_{\text{under the null}}[\ell(X_8,y)|Y_8=y]  \\
	&\quad \iff \Pr_{\text{under any alternative}}(X_8=x|Y_8=y)&\geq& \Pr_{\text{under the null}}(X_8=x|Y_8=y). 
	\end{align*}
	\item Fisher's exact test has power 1 against the alternative of full information. 
	\end{enumerate}
\end{corollary}
Corollary~\ref{cor:fisher} shows that only events related to ``higher success than random'' are more extreme under any violation of the null hypothesis than under the null.

\paragraph{Connection to the maximum entropy, entropic regularization, multinomial logit, and rational inattention literature} Theorem~\ref{thm:shanon_inf}  can be interpreted as a reverse of the Bayesian principle of maximum entropy, which dictates that among all prior distributions that incorporate precise knowledge about data, the one which represents the best the current knowledge is the one with maximal entropy. For instance, given knowledge of the mean, the exponential distribution is the distribution with the largest entropy. Given knowledge of mean and variance, the Gaussian distribution is the distribution with maximum entropy. In contrast, Theorem~\ref{thm:shanon_inf} fixes the entropy and considers the optimizing behaviour of the taster: among all distributions with a given entropy, the taster chooses the one that minimizes their expected prediction loss. In that sense, it is closer to entropic regularization methods at the heart of the optimal transport and machine learning literature \citep[e.g.][]{Wilson1969, Peyre2019}.  Its proof provides another illustration of the deep mathematical connection between entropy and the multinomial logit distribution (or the variational principle and Gibbs measure), as already shown in \cite{Jaynes1957} who reinterpreted statistical mechanics in physics using information theory and the maximum entropy principle, in \cite{Shannon1959} and \cite{Anas1983} in statistical estimation and engineering applications, or more recently in the rational inattention economic literature \citep[e.g.,][]{Matejka2015} where the decision maker's optimal information-processing strategy results in probabilistic choices that follow a logit model. 


\section{Proof of Theorem~\ref{thm:shanon_inf}} 
Suppose one proves another version of Theorem~\ref{thm:shanon_inf} where the inequality constraint is replaced by an equality constraint. Then, using the fourth result of this new version, all results of the original Theorem~\ref{thm:shanon_inf} are straightforward to obtain. Hence, the version with an equality constraint is proved below.


\medskip
\noindent \ref{res:thm1}. $H(P^*_{X_N|Y_N}(h))=h$. Under an equality constraint, this is an immediate consequence of the existence and uniqueness of $P^*_{X_N|Y_N}(h)$ established in Point \ref{res:thm4} below.

\medskip
\noindent \ref{res:thm3}. $P_{X_N|Y_N}^*(\overline h)={\rm Uniform}(\mathcal X_{N,n})$. It suffices to show that for any finite set $\mathcal W$, the collection of distributions $P$ on $\mathcal W$ such that $H(P)=\log(\vert\mathcal W\vert)$ is the singleton $\{{\rm Uniform}(\mathcal W)\}$. This is true since $H$ is uniquely maximized at ${\rm Uniform}(\mathcal W)$ over the collection of distributions $P$ on $\mathcal W$, with maximum value $H({\rm Uniform}(\mathcal W))=\log(\abs{\mathcal W})$. A  proof of this well-known result from information theory is given for completeness. Label $\mathcal W=\{w_1,\ldots,w_{\abs{\mathcal W}}\}$ and, for all $i\in\{1,\ldots,\abs{\mathcal W}\}$, let $p_i=\Pr_{W\sim P}(W=w_i)$, leaving the dependence on $P$ implicit. Note that the mapping $\phi:x\mapsto x\log x$ is strictly convex on $[0,1]$ and the entropy writes $H(P)=-\sum_{i=1}^{\abs{\mathcal W}}\phi(p_i)$. By Jensen's inequality,
\[
\phi\left(\frac{\sum_{i=1}^{\abs{\mathcal W}}p_i}{\abs{\mathcal W}}\right)\leq \frac{\sum_{i=1}^{\abs{\mathcal W}}\phi(p_i)}{\abs{\mathcal W}}=-\frac{H(P)}{\abs{\mathcal W}}.
\]
Hence, $H(P)\leq-\abs{\mathcal W}\phi\left(\frac{\sum_{i=1}^{\abs{\mathcal W}}p_i}{\abs{\mathcal W}}\right)$. From $\sum_{i=1}^{\abs{\mathcal W}}p_i=1$, deduce $H(P)\leq \log(\abs{\mathcal W})$. If $P={\rm Uniform}(\mathcal W)$, then $p_i=1/\abs{\mathcal W}$ so that $H(P)=\log(\abs{\mathcal W})$. That the maximum of $H$ is uniquely attained at ${\rm Uniform}(\mathcal W)$ follows from the fact that the Hessian matrix of $H$ at $P$, denoted $\nabla_P^2 H$, takes value
$$\nabla_P^2 H=\begin{pmatrix}
-1/p_1 & 0 & \cdots & 0 \\ 
 0& -1/p_2 & \ddots & \vdots \\
\vdots & \ddots & \ddots & 0 \\
0& \cdots &0& -1/p_{\abs{\mathcal W}}
\end{pmatrix}=\begin{pmatrix}
-1/\abs{\mathcal W} & 0 & \cdots & 0 \\ 
 0& -1/\abs{\mathcal W} & \ddots & \vdots \\
\vdots & \ddots & \ddots & 0 \\
0& \cdots &0& -1/\abs{\mathcal W}
\end{pmatrix}. $$
This matrix is negative definite, and hence $H$ is strictly concave.

\medskip
\noindent \ref{res:thm4}. {\em For all $h,h'\in(0,\overline h]$ such that $h\leq h'$, $\ell(X_N,Y_N)$ under $P_{X_N|Y_N}^*(h')$ first-order stochastically dominates $\ell(X_N,Y_N)$ under $P_{X_N|Y_N}^*(h)$.} Fix $y\in \mathcal X_{N,n}$ and let us characterize the interior solution of 
\begin{equation}\label{eq:optim_proof}
\begin{array}{cc}
	\min_{P_{X_N|Y_N=y}}\E_{P_{X_N|Y_N=y}}[\ell(X_N,y)] \\
	\text{subject to} \quad - \sum_{x\in\mathcal X_{N,n}}P_{X_N|Y_N=y}(x)\log(P_{X_N|Y_N=y}(x))=h.
\end{array}
\end{equation}
To simplify notation, suppose w.l.o.g.~that $2n=N$. Let $a_k={n\choose n-k+1}{n\choose k-1}$, $b_k=n-k+1$, $\boldsymbol a=(a_1,\ldots,a_{n+1})'$, and $\boldsymbol b=(b_1,\ldots,b_{n+1})'$. The number $a_k$ is the number of events associated with $b_k$ successes out of the first $n$ trials. Rewriting Problem \eqref{eq:optim_proof} as a maximization problem, it is equivalent to finding an interior solution $\boldsymbol p:=(p_1,\ldots, p_{n+1})'>\boldsymbol 0$ that maximizes
$$(\boldsymbol a\odot\boldsymbol b)^\top\boldsymbol p=\sum_{k=1}^{n+1}(n-k+1)a_kp_k$$
subject to
\begin{align*}
	\boldsymbol a^\top\boldsymbol p&=\sum_{k=1}^{n+1}a_kp_k=1, \\
	-\boldsymbol a^\top(\boldsymbol p\odot\log(\boldsymbol p))&=-\sum_{k=1}^{n+1}a_kp_k\log(p_k)=h,
\end{align*}
where $\odot$ denotes the Hadamard product. The Lagrangian associated with this linear program with nonlinear constraints is
\[
\mathcal L(\boldsymbol p,\lambda,\mu)=-(\boldsymbol a\odot\boldsymbol b)^\top\boldsymbol p+\lambda[1-\boldsymbol a^\top\boldsymbol p]+\mu[h+\boldsymbol a^\top(\boldsymbol p\odot\log(\boldsymbol p))].
\]
For $h\in(0,\overline h)$, the Karush-Kuhn-Tucker conditions for an interior solution are
\begin{align}
	\partial \mathcal L/\partial p_k = 0 & \iff-a_kb_k-\lambda a_k+\mu a_k(\log(p_k)+1)=0 \quad \forall k\in\{1,\ldots,n+1\},\label{eq:foc1} \\
	\partial \mathcal L/\partial \lambda = 0 & \iff \boldsymbol a^\top\boldsymbol p=1, \label{eq:foc2} \\
	\partial \mathcal L/\partial \mu = 0 & \iff -\boldsymbol a^\top(\boldsymbol p\odot\log(\boldsymbol p))=h. \label{eq:foc3}
\end{align}
The first-order conditions displayed in Equation~\eqref{eq:foc1} rewrite
\begin{equation}\label{eq:imp1} 
p_k=\exp(b_k/\mu-1)\exp(\lambda/\mu) \quad \forall k\in\{1,\ldots,n+1\}.
\end{equation}
Equation~\eqref{eq:imp1} combined with the first-order condition~\eqref{eq:foc2} implies the logit form
\begin{equation}\label{eq:logit}
p_k=\frac{\exp(b_k/\mu)}{\sum_{\ell=1}^{n+1}a_\ell\exp(b_\ell/\mu)} \quad \forall k\in\{1,\ldots,n+1\}.
\end{equation}
Conclude that $P^*_{X_N|Y_N=y}(h)(x)=\sum_{k=0}^n1\{\ell(x,y)=k\}p_{k+1}$ is unique. The desired first-order stochastic dominance result is equivalent to: for any non-decreasing mapping $u:\R\to\R$,
\begin{equation}\label{eq:key_eq}
	\sum_{k=1}^{n+1}u(b_k)a_kp_k(\mu,h) \geq \sum_{k=1}^{n+1}u(b_k)a_kp_k(\mu',h').
\end{equation}
For $h=\overline h$ or $h=h'$ the result is trivial. Suppose $0< h< h'\leq \overline h$. First, taking the derivative of Equation~\eqref{eq:logit} yields
\begin{equation}\label{eq:prob_deriv}
	\frac{\partial p_k}{\partial \mu}=\frac{p_k}{\mu^2}\left(\frac{\sum_{\ell=1}^{n+1}a_\ell b_\ell \exp(b_\ell/\mu)}{\sum_{\ell=1}^{n+1}a_\ell \exp(b_\ell/\mu)}-b_k\right)=\frac{p_k}{\mu^2}\left(\E_{x\sim P^*_{X_N|Y_N=y}(h)}[N-\ell(x,y)]-b_k\right).
\end{equation}
Second, let us show that ${\rm d} \mu/{\rm d} h>0$. First-order condition~\eqref{eq:foc3} writes
\[
\underbrace{\frac{1}{{\sum_{\ell=1}^{n+1}a_\ell\exp(b_\ell/\mu)}}\sum_{k=1}^{n+1}a_k\exp(b_k/\mu)\left[-b_k/\mu+\log\left(\sum_{\ell=1}^{n+1}a_\ell\exp(b_\ell/\mu)\right)\right]}_{=:f(\mu)}=h.
\]
Taking the derivative of $f$ yields
\begin{align*}
	 f'(\mu)  &= \Big( \Big[\sum_{k=1}^{n+1}\left(\frac{-a_kb_k}{\mu^2}\right)\exp(b_k/\mu)\Big\{-b_k/\mu+\log\left(\sum_{\ell=1}^{n+1}a_\ell\exp(b_\ell/\mu)\right)\Big\} \\
	&\quad +a_k\exp(b_k/\mu)\Big\{b_k/\mu^2+\frac{\sum_{\ell=1}^{n+1}\left(\frac{-a_\ell b_\ell}{\mu^2}\right)\exp(b_\ell/\mu)}{\sum_{\ell=1}^{n+1}a_\ell\exp(b_\ell/\mu)}\Big\} \Big]\Big[\sum_{\ell=1}^{n+1}a_\ell \exp(b_\ell/\mu)\Big] \\
	& \quad + \Big[\sum_{\ell=1}^{n+1}\frac{a_\ell b_\ell}{\mu^2} \exp(b_\ell/\mu)\Big] \Big[\sum_{k=1}^{n+1}a_k\exp(b_k/\mu)\Big\{-b_k/\mu+\log\left(\sum_{\ell=1}^{n+1}a_\ell\exp(b_\ell/\mu)\right)\Big\}\Big]\Big) \\
	&\quad \Big/  \Big(\sum_{\ell=1}^{n+1}a_\ell \exp(b_\ell/\mu)\Big)^2 \\
	& = \Big( \Big[\sum_{k=1}^{n+1}\left(\frac{a_kb_k^2}{\mu^3}\right)\exp(b_k/\mu)\\
	&\quad +a_k\exp(b_k/\mu)\Big\{b_k/\mu^2+\frac{\sum_{\ell=1}^{n+1}\left(\frac{-a_\ell b_\ell}{\mu^2}\right)\exp(b_\ell/\mu)}{\sum_{\ell=1}^{n+1}a_\ell\exp(b_\ell/\mu)}\Big\} \Big]\Big[\sum_{\ell=1}^{n+1}a_\ell \exp(b_\ell/\mu)\Big] \\
	& \quad + \Big[\sum_{\ell=1}^{n+1}\frac{a_\ell b_\ell}{\mu^2} \exp(b_\ell/\mu)\Big] \Big[\sum_{k=1}^{n+1}\left(\frac{-a_kb_k}{\mu}\right)\exp(b_k/\mu)\Big]\Big) \Big/  \Big(\sum_{\ell=1}^{n+1}a_\ell \exp(b_\ell/\mu)\Big)^2.
\end{align*}
This expression can be reorganized as  
\begin{align*}
	f'(\mu)&  = \Big( \Big[\sum_{k=1}^{n+1}\left(\frac{a_kb_k^2}{\mu^3}\right)\exp(b_k/\mu) +a_k\exp(b_k/\mu)\Big\{\frac{\sum_{\ell=1}^{n+1}\left(\frac{-a_\ell b_\ell}{\mu^2}\right)\exp(b_\ell/\mu)}{\sum_{\ell=1}^{n+1}a_\ell\exp(b_\ell/\mu)}\Big\} \Big]\\
	&\quad \times \Big[\sum_{\ell=1}^{n+1}a_\ell \exp(b_\ell/\mu)\Big]+\Big[\sum_{\ell=1}^{n+1}\frac{a_\ell b_\ell}{\mu^2} \exp(b_\ell/\mu)\Big] \Big[\sum_{k=1}^{n+1}a_k\exp(b_k/\mu)(1-b_k/\mu)\Big]\Big) \\
	&\quad \Big/  \Big(\sum_{\ell=1}^{n+1}a_\ell \exp(b_\ell/\mu)\Big)^2. 
\end{align*}
This expression further simplifies to 
\begin{align*}
 f'(\mu) & = \Big( \Big[\sum_{k=1}^{n+1} \frac{a_kb_k^2}{\mu^3}\exp(b_k/\mu)\Big]\Big[\sum_{\ell=1}^{n+1}a_\ell \exp(b_\ell/\mu)\Big] \\
	&\quad  +\Big[\sum_{k=1}^{n+1}a_k\exp(b_k/\mu)\Big]\Big[\sum_{\ell=1}^{n+1}\left(\frac{-a_\ell b_\ell}{\mu^2}\right)\exp(b_\ell/\mu)\Big] \Big) \\
	&\quad +\Big[\sum_{\ell=1}^{n+1}\frac{a_\ell b_\ell}{\mu^2} \exp(b_\ell/\mu)\Big] \Big[\sum_{k=1}^{n+1}a_k\exp(b_k/\mu)(1-b_k/\mu)\Big]\Big) \\
	&\quad \Big/  \Big(\sum_{\ell=1}^{n+1}a_\ell \exp(b_\ell/\mu)\Big)^2 \\
	&  = \Big( \Big[\sum_{k=1}^{n+1} \frac{a_kb_k^2}{\mu^3}\exp(b_k/\mu)\Big]\Big[\sum_{\ell=1}^{n+1}a_\ell \exp(b_\ell/\mu)\Big] \\
	&\quad -\Big[\sum_{\ell=1}^{n+1}\frac{a_\ell b_\ell}{\mu^2} \exp(b_\ell/\mu)\Big] \Big[\sum_{k=1}^{n+1}\frac{a_kb_k}{\mu}\exp(b_k/\mu)\Big]\Big)  \Big/  \Big(\sum_{\ell=1}^{n+1}a_\ell \exp(b_\ell/\mu)\Big)^2.
\end{align*}
Since $\mu>0$, the sign of $f'(\mu)$ is the same as the sign of 
\[
\underbrace{\Big[\sum_{k=1}^{n+1} a_kb_k^2\exp(b_k/\mu)\Big]\Big[\sum_{\ell=1}^{n+1}a_\ell \exp(b_\ell/\mu)\Big]}_{=:A(\mu)} -\underbrace{\Big[\sum_{\ell=1}^{n+1}a_\ell b_\ell\exp(b_\ell/\mu)\Big]^2}_{=:B(\mu)}.
\]
By developing products and using mathematical induction on $n\in\mathbb N$, it can be shown that, for all $\mu\in\R$,
\begin{align*}
	A(\mu)-B(\mu)&=\sum_{k=1}^{n+1}\sum_{\ell=1}^{n+1}a_ka_\ell\exp(b_k/\mu)\exp(b_\ell/\mu) b_k(b_k-b_\ell)\geq 0.
\end{align*}
Conclude
\[
\frac{{\rm d} \mu}{{\rm d} h}=[f'(\mu)]^{-1}> 0.
\]
Next, because $b_1>b_2>\cdots>b_{n+1}$ and $\E_{x\sim P^*_{X_N|Y_N=y}(h)}[N-\ell(x,y)]\in(b_{n+1},b_1)$, Equation~\eqref{eq:prob_deriv} implies that there exists some $\bar k\in\{1,\ldots,n+1\}$ such that
\[
\left\{
\begin{array}{ll}
	p_k(\mu,h)\geq p_k(\mu',h') \quad \forall k\leq \bar k, \\
	p_k(\mu,h)\leq p_k(\mu',h') \quad \forall k> \bar k.
\end{array}\right.
\]
Again, since $b_1>b_2>\cdots>b_{n+1}$ and $u$ is non-decreasing, the decomposition 
\begin{align*}
	& \sum_{k=1}^{n+1}u(b_k)a_kp_k(\mu,h)\\
	&\quad  =\sum_{k=1}^{n+1}u(b_k)a_kp_k(\mu',h')+\sum_{k=1}^{n+1}u(b_k)a_k[p_k(\mu,h)-p_k(\mu',h')] \\
	& \quad=\sum_{k=1}^{n+1}u(b_k)a_kp_k(\mu',h') +\sum_{k\leq \bar k}u(b_k)\underbrace{a_k[p_k(\mu,h)-p_k(\mu',h')}_{\geq 0} + \sum_{k> \bar k}u(b_k)\underbrace{a_k[p_k(\mu,h)-p_k(\mu',h')}_{\leq 0} \\
	&\quad \geq \sum_{k=1}^{n+1}u(b_k)a_kp_k(\mu',h') +u(b_{\bar k})\sum_{k\leq \bar k}a_k[p_k(\mu,h)-p_k(\mu',h')] +u(b_{\bar k})\sum_{k> \bar k}a_k[p_k(\mu,h)-p_k(\mu',h')] \\
	& \quad = \sum_{k=1}^{n+1}u(b_k)a_kp_k(\mu',h')+u(b_{\bar k})\left[\underbrace{\sum_{k=1}^{n+1}a_kp_k(\mu,h)}_{=1}-\underbrace{\sum_{k=1}^{n+1}a_kp_k(\mu',h')}_{=1}\right] \\
	& \quad = \sum_{k=1}^{n+1}u(b_k)a_kp_k(\mu',h')
\end{align*}
completes the proof of  \eqref{eq:key_eq}.

\medskip
\noindent \ref{res:thm2}.  $\lim_{h\to0^+}\sup_{x\in\mathcal X_{N,n}}\abs{P_{X_N|Y_N}^*(h)(x)-\delta_{Y_N}(x)}=0$. Since $\mu>0$ and $\frac{{\rm d} \mu}{{\rm d} h}> 0$, the result follows by taking $\lim_{\mu\to0^+} p_k(\mu)=1\{k=1\}/a_1=1\{k=1\}$ in Equation~\eqref{eq:logit}.

\medskip
\noindent The proof of Theorem~\ref{thm:shanon_inf} is complete.

\bibliographystyle{agsm} 

\bibliography{biblio}
\end{document}